 \documentclass[preprint2]{aastex6}
 \usepackage{natbib}

 \usepackage{graphicx}
 \usepackage{float}
 \usepackage{csquotes} 
 \usepackage{gensymb}

\usepackage[normalem]{ulem}
\usepackage{array}
\newcolumntype{L}[1]{>{\raggedright\let\newline\\\arraybackslash\hspace{0pt}}m{#1}}
\newcolumntype{C}[1]{>{\centering\let\newline\\\arraybackslash\hspace{0pt}}m{#1}}
\newcolumntype{R}[1]{>{\raggedleft\let\newline\\\arraybackslash\hspace{0pt}}m{#1}}

\usepackage{color}

\fullcollaborationName{The Friends of AASTeX Collaboration}

\begin{document}


\title{\textit{Cloud Atlas}: DISCOVERY OF PATCHY CLOUDS AND HIGH-AMPLITUDE ROTATIONAL MODULATIONS IN A YOUNG, EXTREMELY RED L-TYPE BROWN DWARF}


\author{Ben W. P. Lew\altaffilmark{1}, Daniel Apai\altaffilmark{1,2,3},  Yifan Zhou\altaffilmark{2}, Glenn Schneider\altaffilmark{2}, Adam J. Burgasser\altaffilmark{4}, Theodora Karalidi\altaffilmark{2}, Hao Yang\altaffilmark{2}, Mark S. Marley\altaffilmark{5}, Nicolas B. Cowan\altaffilmark{6}, Luigi R. Bedin\altaffilmark{7}, Stanimir A. Metchev\altaffilmark{8}, Jacqueline Radigan\altaffilmark{9}, Patrick J. Lowrance\altaffilmark{10}}



\altaffiltext{1}{Department of Planetary Science/Lunar and Planetary Laboratory, The University of Arizona, 1640 E. University Blvd.,
Tucson, AZ 85718, USA; wplew@lpl.arizona.edu}
\altaffiltext{2}{Department of Astronomy/Steward Observatory, The University of Arizona, 933 N. Cherry Ave., Tucson, AZ, 85721, USA}
\altaffiltext{3}{Earths in Other Solar Systems Team, NASA Nexus for Exoplanet System Science}
\altaffiltext{4}{Center for Astrophysics and Space Science, University of California San Diego, La Jolla, CA 92093, USA}
\altaffiltext{5}{NASA Ames Research Center, Naval Air Station, Moffett Field, Mountain View, CA 94035, USA} 
\altaffiltext{6}{Department of Physics \& Astronomy, Amherst College, Amherst, MA 01002, USA}
\altaffiltext{7}{INAF - Osservatorio Astronomico di Padova, Vicolo dell'Osservatorio 5, I-35122 Padova, Italy}
\altaffiltext{8}{Department of Physics and Astronomy, The University of Western Ontario, 1151 Richmond Avenue, London, ON N6A 3K7, Canada} 
\altaffiltext{9}{Department of Physics, Utah Valley University, 800 West University Parkway, Orem, UT 84058, USA} 
\altaffiltext{10}{Infrared Processing and Analysis Center, MS 100-22, California Institute of Technology, Pasadena, CA 91125, USA}

\begin{abstract}
Condensate clouds fundamentally impact the atmospheric structure and spectra of exoplanets and brown dwarfs but the connections between surface gravity, cloud structure, dust in the upper atmosphere, and the red colors of some brown dwarfs remain poorly understood. Rotational modulations enable the study of different clouds in the same atmosphere, thereby providing a method to isolate the effects of clouds. Here we present the discovery of high peak-to-peak amplitude (8\%)  rotational modulations in a low-gravity, extremely red (J-$K_s$=2.55) L6 dwarf WISEP J004701.06+680352.1 (W0047). Using the \textit{Hubble Space Telescope} (HST) time-resolved grism spectroscopy we find a best-fit rotational period (13.20$\pm0.14$ hours) with a larger amplitude at 1.1 micron than at 1.7 micron.  This is the third largest near-infrared variability amplitude measured in a brown dwarf, demonstrating that large-amplitude variations are not limited to the L/T transition but are present in some extremely red L-type dwarfs. We report a tentative trend between the wavelength dependence of relative amplitude, possibly proxy for small dust grains lofted in the upper atmosphere, and the likelihood of large-amplitude variability. By assuming forsterite as haze particle, we successfully explain the wavelength dependent amplitude with submicron-sized haze particles sizes of around 0.4 $\mu m$. W0047 links the earlier spectral and later spectral type brown dwarfs in which rotational modulations have been observed; the large amplitude variations in this object make this a benchmark brown dwarf for the study of cloud properties close to the L/T transition.  
\end{abstract}
\keywords{brown dwarfs --- stars: atmospheres --- stars: individual: WISEP J004701.06+680352.1 --- stars: low-mass}



\section{Introduction} \label{sec:intro}
Condensate clouds influence the colors, spectra, atmospheric structure, and evolution of brown dwarf and giant exoplanets. The physics of grain formation and cloud evolution are perhaps the least understood major factors in ultracool atmosphere models. Large numbers of comparative spectroscopic studies \citep[e.g.,][]{burgasser2014} and variability surveys \citep[e.g.,][]{metchev2015, radigan2014, buenzli2014,wilson2014, heinze2015} of brown dwarfs have provided important insights into the properties and effects of clouds. Each brown dwarf, however, has a unique set of the fundamental atmospheric parameters (e.g., composition, mass, surface gravity, age, cloud properties, temperature, vertical mixing and perhaps even rotation), whose effect on the spectra are degenerate. This has made it difficult to isolate the effects of clouds from those of other parameters \citep[e.g.,][]{looper2008,stephens2009}. Atmospheric models incorporating condensate cloud structures predicted observable brightness variations in rotating objects if the cloud properties are heterogeneous \citep[e.g.,][]{ackerman2001} and motivated an almost decade-long search for varying brown dwarfs \citep[e.g.,][]{martin2001,clarke2002,koen2003, bailer2003,maiti2005,morales2006,littlefair2008, goldman2008, clarke2008, bailer2008}. Modern systematic infrared precision surveys resulted in the first large-amplitude variable brown dwarf detections \citep[e.g.,][]{artigau2009,radigan2012,biller2015} and recent rotational phase studies have enabled using these objects as probes of different cloud structures. In particular, HST near-infrared time-resolved spectroscopy has revealed that the rotational phase modulation's amplitude has no or only mild wavelength dependence \citep[][]{apai2013,buenzli2015ab,yang2015}. The observed photometric and spectroscopic modulations in L/T transition objects have been explained with cloud thickness and temperature variations \citep[][]{artigau2009,radigan2012, apai2013,burgasser2013,biller2013}; and in L5-dwarfs evidence was found for high-altitude hazes \citep[][]{yang2015}. A key difference between the variability seen in L/T and L dwarfs is the significant reduction (2x) in the 1.4 micron water-band variability amplitude in L/T dwarfs and not in L-dwarfs, a probable indicator of high-altitude dust in L-dwarfs \citep{yang2015}. These studies showed that, in essence, by measuring the amplitudes in and out of the water band it is possible to determine the pressure at which clouds reside.

The extremely red colors seen in a few brown dwarfs (both young and older) and in some directly imaged exoplanets suggests that these objects may have exceptionally dusty atmospheres \citep[e.g.,][]{knapp2004,looper2008, barman2011,marocco2014, hiranaka2016, allers2016}.  Although an attractive possibility, non-equilibrium chemistry \citep[e.g.,][]{saumon2003,tremblin2016}, metallicity \citep{looper2008} and other effects may also play a role in these unusual atmospheres. The apparently diverse age \citep[e.g.,][]{kirkpatrick2010} of these extremely red objects also raises questions on the origins of the purportedly dusty atmospheres.

Our Hubble Space Telescope (HST) Large Treasury program \textit{Cloud Atlas} (PI: D. Apai, GO14241) is studying the impact of surface gravity on the vertical structure of condensate clouds via rotational phase mapping of low ($log~g \leqslant 4.0$)  and intermediate-gravity ($4.0 < log~g <  5.0 $) brown dwarfs and exoplanets. The observations are divided into two phases for each target: a 2-orbit {\em Variability Amplitude Assessment Survey} (VAAS) and follow-on with detection, 6-orbit {\em Deep Look Observations} (DLO). The results of the VAAS observations are used to prioritize targets for DLO studies.

Here we present the first results from the \textit{Cloud Atlas} program, the discovery of high peak-to-peak amplitude ($>$8\%) rotational modulations in an extremely red and young 20~M$_{Jup}$ brown dwarf. In this Letter we first discuss the properties of our target, followed by a brief summary of the data reduction and analysis, our key results, and a concise discussion of the relevance of the discovery. 

\subsection{WISEP J004701.06+680352.1}

WISEP J004701.06$+$680352.1 (hereafter W0047) was discovered by \citet{gizis2012} as an extremely red (2MASS J-$K_{s}$= 2.55)  L-type brown dwarf. Following the  \citet{allers2013a}'s spectroscopic classification system \citet{gizis2015} (hereafter G15)  identified it as a L6.5-type intermediate-gravity brown dwarf. Based on comparison to state-of-the art atmospheric models \citep[][]{ackerman2001,marley2002,madhusudhan2011,tsuji2002} \citet{gizis2012} explained the extremely red color of W0047 due to an exceptionally thick atmospheric cloud layer. G15 also measured its parallactic distance (d = 8.06$\pm$0.04~pc) and proper motion, concluding that it is associated with AB Dor moving group with 99.96\% probability according to BANYAN II model of the solar neighborhood \citep{gagne2014}. This moving group membership constrains the age of W0047 to $\sim100-125$~Myr \citep{luhman2005,barenfeld2013}. G15 also found a radial velocity of v $\sin$ i = 4.3$\pm 2.2 $km/s. With a well-determined age, parallax, and high-quality spectra W0047 is among the best-characterized nearby brown dwarfs; comparing its luminosity and temperature to evolutionary models by \citet{burrows1997}  and \citet{Chabrier2000} G15 also estimated the mass of W0047 to be 18--20$M_{Jup}$ and its radius to be 1.17--1.24$ R_{Jup}$.

As one of the reddest known L dwarfs (e.g., PSO J318.5338-22.8603 \citep{liu2013}, 2M1207b \citep{chauvin2004}, WISEA J114724.10-204021.3 \citep{schneider2016}, 2MASS J11193254-1137466 \citep{kellogg2015}, ULAS J222711-004547 \citep{marocco2014}), W0047's young age with intermediate surface gravity properties make this a particularly exciting target for the study of the connections between surface gravity and dusty atmospheres.

\section{Observations and Data Reduction}

W0047 is the first DLO target observed in the \textit{Cloud Atlas} program. Analysis of the VAAS data showed a statistically significant and strong flux modulation over the 2 HST orbits of the data set, based on which we flagged W0047 as a probable variable brown dwarf and triggered 6-orbit long DLO observations. All VAAS data will be presented in an upcoming paper (Apai et al., in prep.) and here we only focus on the more detailed DLO dataset. 

W0047 was observed with HST WFC3's G141 grism  \citep{mackenty2010}, which covers the wavelength range between 1.075 $\mu m$ and 1.7 $\mu m$ with a resolving power of $\sim$ 130 at 1.4 $\mu m$. The 6 orbits were executed consecutively on 2016 June 6th (Visit 2, DLO, observation set ID: H1). For each orbit at least one direct image was obtained in the F132 N filter to determine our target's precise position on the detector for accurate spectral extraction. 11 subarray (256 $\times$ 256 pixels) grism images were obtained in the SPARS25 sampling mode with total 201.4~s exposure time during each orbit. 

We followed the same data reduction procedure as discussed in our group's previous papers \citep[][]{apai2013, buenzli2014, yang2015}, with the following differences: \\
1. We linearly interpolated all data quality flagged pixels with good neighbor pixels at same row except for cosmic-ray affected (DQ bit 13 = 1) pixels and when consecutive three or more bad pixels appeared in the same row. Cosmic ray affected pixels were corrected as in \citet{buenzli2014}. Any spectral point ($F_{\lambda}$) that was calculated from uncorrected bad pixel was interpolated by adjacent spectral points.\\
2. We selected a source-free region in the images to measure the sky background to scale the sky image template and subtracted from the images.

The major systematics observed in the first two orbits is the detector's ramp effect (see \citealt{berta2012}). This was corrected by fitting the light curve of nearby, similarly bright ($\Delta mag$ = 0.68 dimmer than W0047 in F132N filter) comparison star to an exponential function. At wavelength longer than 1.5 $\mu m$, the spectrum of the comparison star is out of the field of view so ramp correction was applied by fitting the light curve of W0047 to an exponential function for the narrow H and $\mathrm{CH_4}$ \& $\mathrm{H_2O}$ narrow band light curves.

Using the 2MASS relative spectral response curves from \citet{cohen2003}, we integrated our spectra to calculate the W0047's 2MASS J and modified H band, which is identical to the 2MASS H-band filter, but is truncated at 1.7 $\mu m$.

\section{Results}

\subsection{Periodic modulations}
\label{sec:period}
\begin{figure*}
\centering
\includegraphics[scale=0.4]{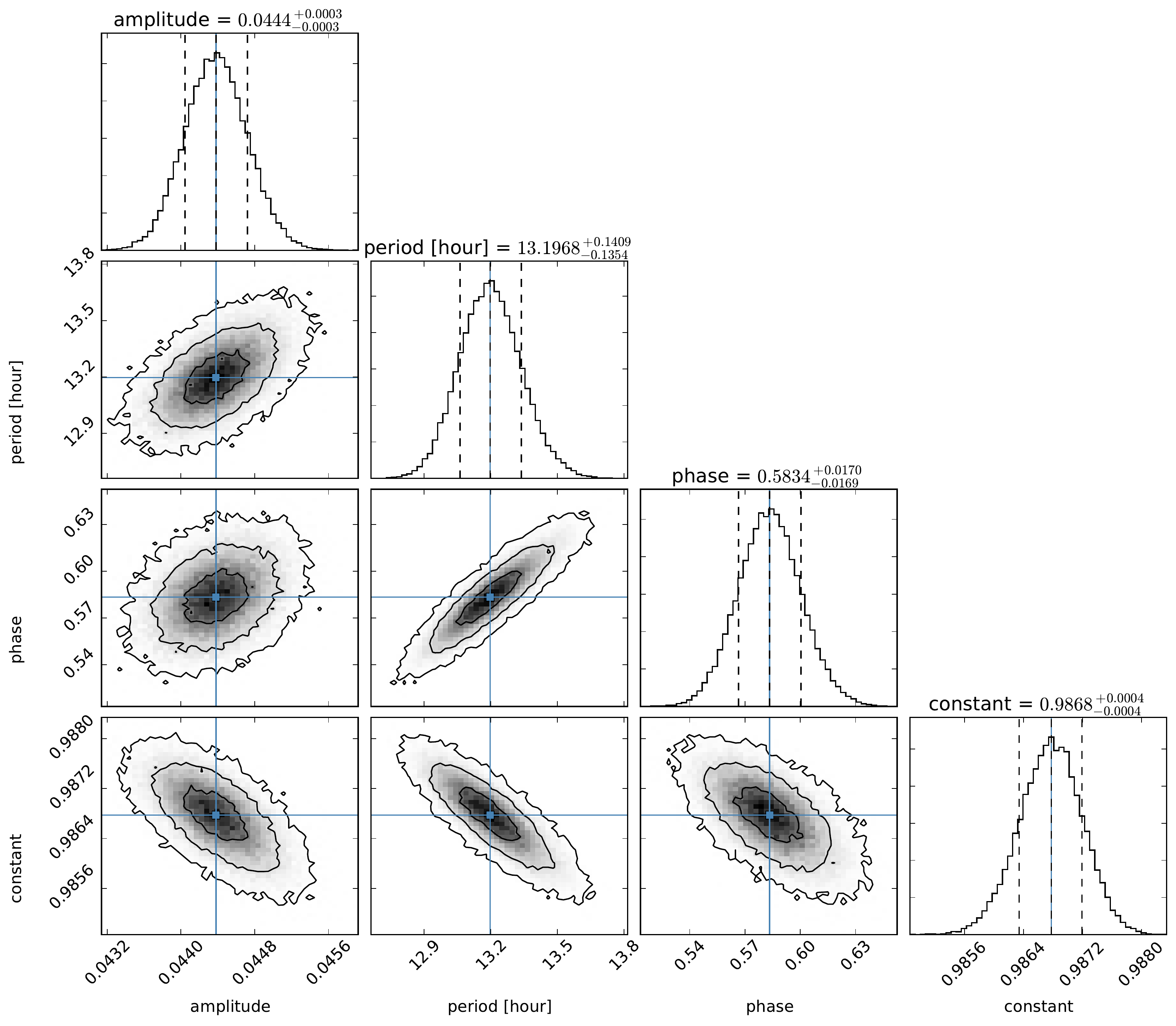}
\caption{The posterior distribution from MCMC fit with uniform prior for a sine curve for all DLO orbits.The vertical dashed lines are the 16 and 85 percentiles respectively. Blue lines indicate the median of posterior distribution. The contour lines show the 1-$\sigma$, 2-$\sigma$ and 3-$\sigma$ level. The amplitude is the amplitude of white light curve flux which is normalized by mean flux of orbit-3. Phase is measured from the beginning of DLO orbit.}
\label{fig:mcmc}
\end{figure*}
The ramp-effect corrected light curve shows a complex lightcurve with a nearly sine-wave-like carrier. In a paper in preparation we are exploring the complex light curve shape, but - as the lightcurve is dominated by a sinusoidal carrier - for the purposes of this discovery letter we fit the light curve with a sine wave. The residual of best sine wave fit is less than 1.1\% of the white lightcurve's normalized flux (see upper left panel of Fig. \ref{fig:phaselightcurve}) and is thus a good approximation of the variations. We fit the light curve with a Markov Chain Monte Carlo method \citep{lmfit}, which also allows us to quantify the uncertainties and degeneracies of the fit. We find a rotational period of 13.20$\pm0.14$ hours (see Fig \ref{fig:mcmc} for the posterior probability distributions). Least $\chi^2$ fit gives period of 13.41$\pm0.25$ hours with reduced $\chi^2 $ 2.76 . We note that our result is based on sinusoidal wave assumption; we cannot exclude a double-peaked light curve due to incomplete phase coverage.

\subsection{Spectral Variations}
\begin{figure*}[!h]
\includegraphics[scale=0.55]{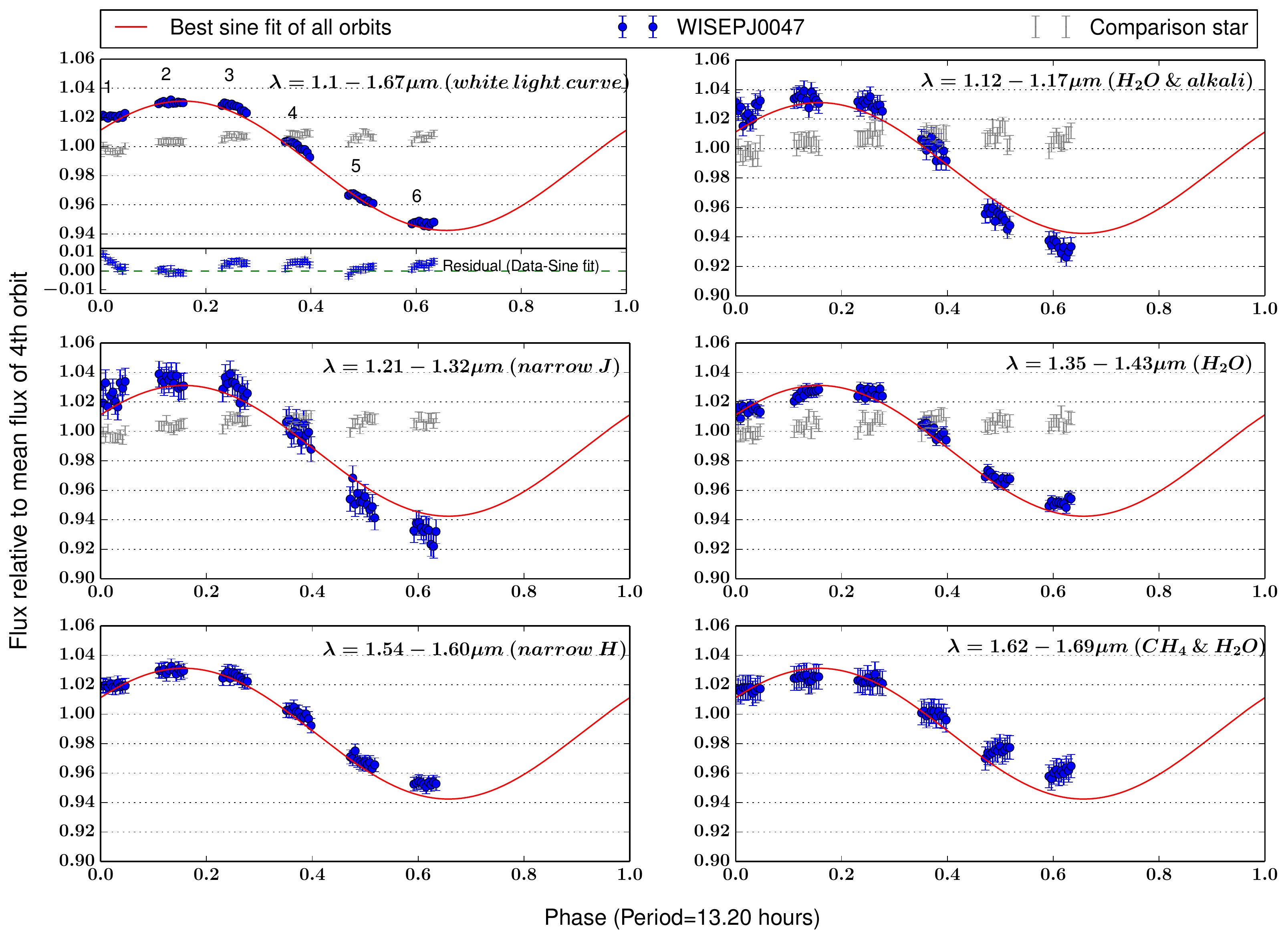}
\caption{ The phase-folded white light curve in comparison with other 5 different narrow band light curves, following \citet{buenzli2012}'s method to probe different atmospheric pressure levels. All variation is in phase and the light curves of shorter wavelengths have slightly larger amplitude. The integrated flux of W0047 of 6 DLO orbits are plotted in blue solid dots, while the comparison star are plotted in grey. The red line show the best sine fit of all orbits with period = 13.20 hours. Orbit numbers are labelled in the white light curve panel (upper left). All light curves are normalized to the mean flux of orbit-4. The spectrum of the comparison star is partly out of view,  so the white light curve of comparison star consists of flux from 1.1 to 1.5 $\mu m$ only and flux in narrow H and $\mathrm{CH_4}$ \& $\mathrm{H_2O}$ narrow band are out of view.}
\label{fig:phaselightcurve}

\end{figure*}

As shown in Figure \ref{fig:phaselightcurve}, we observed a sinusoidal-like variation with a 4\% amplitude (8\% peak-to-peak amplitude). The trend of the white light curve ($1.1 \mu m < \lambda < 1.67 \mu m$) is in phase with the other 5 narrow bands ($\mathrm{H_2O}$ \& alkali, narrow J, $\mathrm{H_2O}$, narrow H, $\mathrm{CH_4}$ \& $\mathrm{H_2O}$) light curves, which probe different atmospheric pressure levels \citep{buenzli2012}. In particular, the white light curve demonstrates that the variability amplitude is much greater than the noise level ($\sim 0.1\%$, signal-to-noise ratio $>$ 600) while in the spectrally extracted narrow bands the noise level is at most around 1\% relative to the mean flux. For a quantitative evaluation of the pressure levels probed by these filters, see \citet{yang2016}.

\subsection{Color Variations}
\begin{figure}[!h]
\flushleft
\includegraphics[width=.52\textwidth]{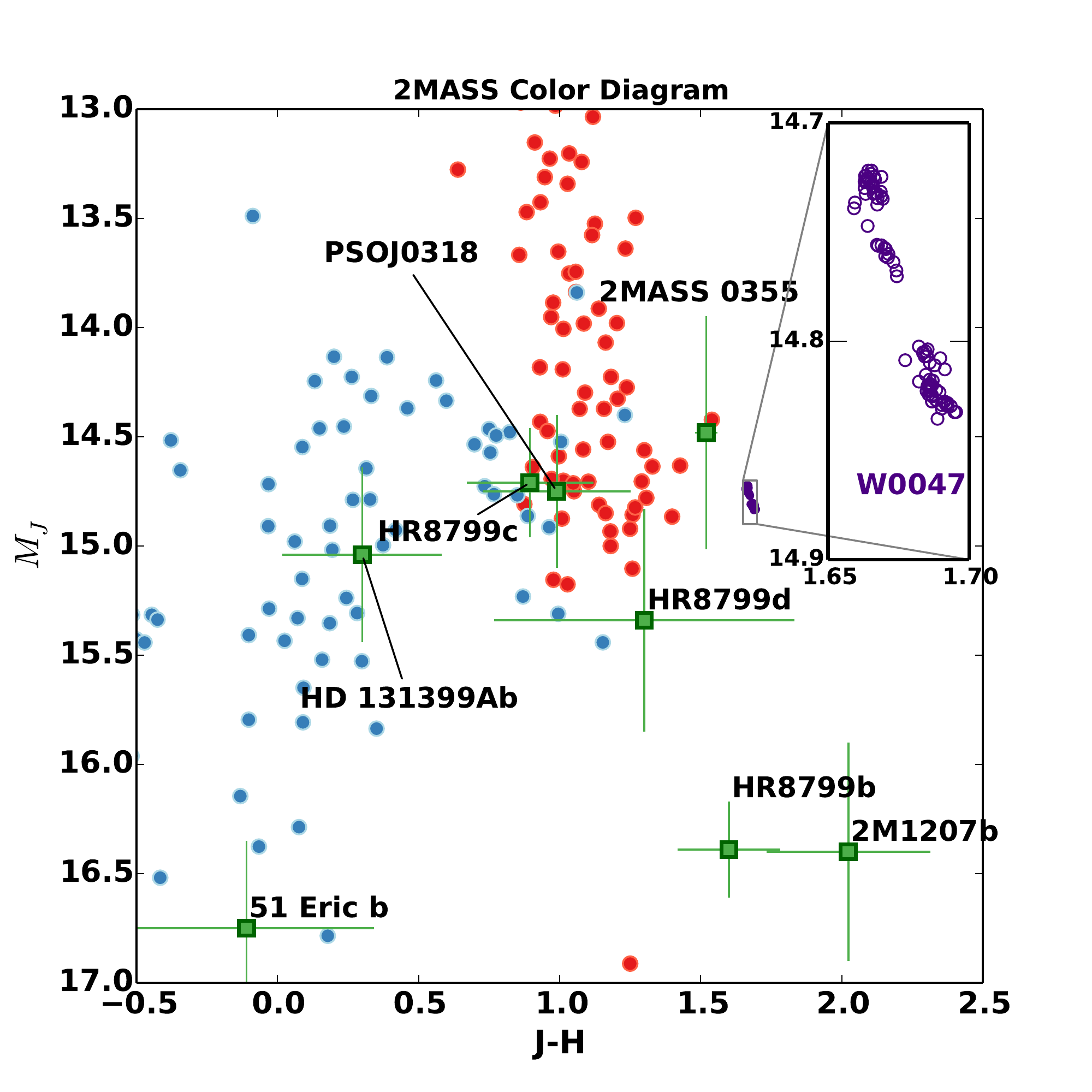}
\caption{The color-magnitude variations of W0047 (purple) vary in {\em parallel} with the L dwarf sequence. L dwarfs and T dwarfs are plotted in red and blue color, respectively.  AB Dor group member (2MASS 0355+1133, or 2MASS 0355) and planetary-mass object (PSOJ0318, HR8799bcd, 2M1207b) are labelled as well. Data are adopted from \citealt{dupuy2012}; for PSOJ0318 from \citet{liu2013}; and for exoplanets \object{HD131399Ab} \citet[][]{wagner2016}, 51 Eri b \citet[][]{macintosh2015}, 2M1207b \citet[][]{chauvin2005,zhou2016}. }
\label{fig:cmd}

\end{figure}
We compare the color variations of W0047 to the color distribution of ultracool dwarfs in a near-infrared color-magnitude diagram (Fig.~\ref{fig:cmd}). 
In figure \ref{fig:cmd}, we used the 2MASS catalog's J and H band value as the mean J and H band value of W0047. The modified H band variability relative to mean modified H band is assumed to be the same as 2MASS H band variability relative to mean 2MASS H band. This assumption is valid as extrapolation of the wavelength-dependent amplitude trend (see middle left of Fig. \ref{fig:ampratio}) shows that the variability (max/min flux ratio) at 1.8 $\mu m$ (the maximum wavelength of 2MASS H band) is less than 1.2\% than the variability at 1.67 $\mu m$. The trajectory of W0047's color variation appears to be qualitatively parallel with the L-dwarf sequence, similarly to what was seen for several T2 dwarfs (e.g., \citealt{apai2013}). 

\subsection{Wavelength-dependent Amplitude}
\begin{figure*}[h]
 	\includegraphics[scale=0.53]{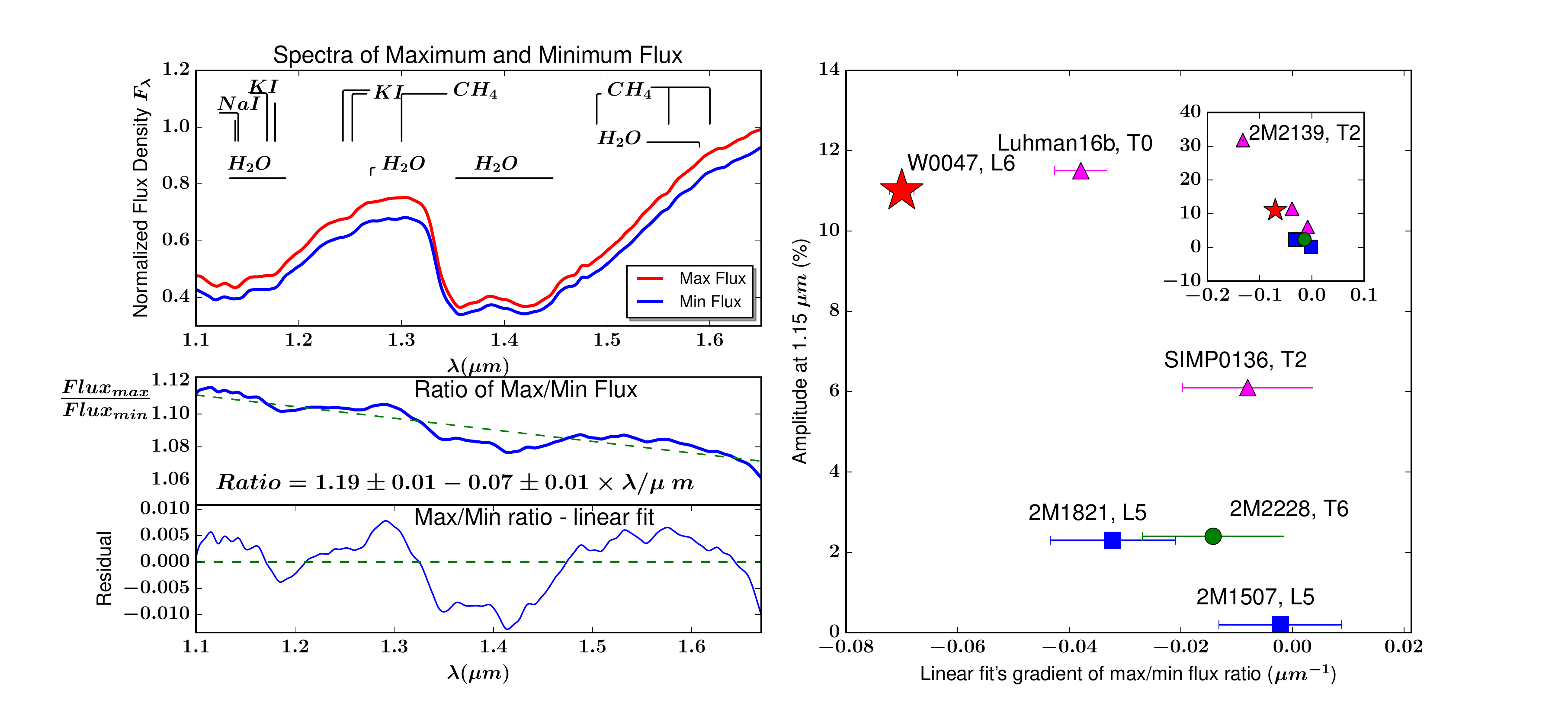}
  	\caption{ \textit{ \bf Upper Left}: Comparison between the mean of 3 brightest spectra (\textit{Red}) and mean of 3 dimmest spectra (\textit{Blue}) of W0047. \textit{\bf Middle Left}: The relative amplitude over $1.1\mu m$ to 1.67$\mu m$, which is the ratio of the mean brightest and dimmest spectra. The relative amplitude is almost linearly decreasing with increasing wavelength, similar to other L dwarfs seen in \citet{yang2015} but different with L/T transition which have lower relative amplitude in water band. The spectra are smoothed by gaussian convolution and the relative amplitude are smoothed by moving average. \textit{\bf Lower Left}: The residual of straight line fit.
	 \textit{\bf Right}: A possible trend between amplitudes at 1.1$\mu m$ and wavelength dependence of amplitude of variable L and T dwarfs that observed with HST. The amplitude is proportional to absolute value of amplitude gradient. At similar amplitude levels, L dwarfs and late T dwarfs are seen to have relatively larger amplitude gradient than L/T transition brown dwarfs. L dwarfs, L/T transition dwarfs and late T dwarfs are marked as blue squares, magenta triangles and green circles respectively. W0047 is marked as red star. Data of 2M1821, 2M1507, SIMP0136, 2M2228 are from \citet{yang2015}; Luhman 16B is from \citet{buenzli2015ab}; 2M2139 is from \citet{apai2013}.}
	 \label{fig:ampratio}

\end{figure*}
In the middle left panel of Fig.~\ref{fig:ampratio} we plot the relative amplitude of the rotational modulations as a function of wavelength, as derived from the ratio of the spectra at the brightest and faintest states (upper left panel), following \citet[][]{apai2013}. The relative amplitude is the highest (11\%) at the shortest wavelength ($\sim1.1~\mu$m) and lowest ($\sim$6.5\%) at the longest wavelength (close to 1.7~$\mu m$). In between these two extremes there is a nearly monotonic trend. A linear regression results in a slope of $\sim$0.07\% / $\mu m$. After subtracting the relative amplitude with the straight-line fit we find that the relative amplitudes are slightly modulated (decrease by 1\% than the linear fit) in the 1.4~$\mu m$ water band.

\subsection{Inclination}
We re-evaluated the $v \sin i$ measurement based on Keck/NIRSPEC data of G15 with \citet{burgasser2016}'s method, leading to a value consistent with G15 but more accurate $v \sin i$ = $6.7^{+0.7}_{-1.4} $km/s. Using radius ($R=1.17 R_{Jup}$) predicted by evolutionary model as discussed in G15, we constrain the inclination of W0047 to $i\sim  33_{-8}^{+5} ~\degree$.

\section{ Discussion}

Our observations find a near-infrared relative amplitude of 11\% for W0047 at the shortest wavelength of our observations (1.1~$\mu m$), making it the third largest amplitude varying brown dwarf (after 2M2139 and before SIMP0136 and Luhman 16b, e.g., \citealt[][]{radigan2012,apai2013,buenzli2015ab}). This discovery demonstrates that high-amplitude near-infrared variability is present not only in L/T transition brown dwarfs \citep[c.f.,][]{radigan2012}, although it may be most common in those transition objects. 

Our analysis of the relative amplitude as a function of wavelength (see Fig. \ref{fig:ampratio}) shows that the relative amplitude of the rotational modulations has a relatively strong color gradient, with the relative amplitude varying by nearly a factor of two (11\% and 7\%) between 1.1 and 1.7~$\mu m$. The factor of two difference in the 1.1$-$1.7 $\mu m$ is the second largest amplitude difference observed in the small set of brown dwarfs that have been studied by HST \citep[][]{buenzli2012,apai2013,buenzli2015ab,yang2015}. In the right panel of Fig. \ref{fig:ampratio} we explore the presence of a possible trend between the relative amplitude at 1.15$\mu m$ and the wavelength dependence of relative amplitude, showing that W0047 appears to be in between L and L/T transition brown dwarfs. This finding is also consistent with the observations by \citet{heinze2015} that identified higher-amplitude variations for T-dwarfs at optical wavelengths. 

We also compare the relative amplitude to those observed in L/T transition brown dwarfs and in mid-L-type brown dwarfs. While all L/T transition brown dwarfs studied until now show a reduced rotational modulation in the 1.4~$\mu m$ water band \citep[e.g.,][]{apai2013,buenzli2015ab}, the two mid-L-type brown dwarfs observed with HST for full rotations show small or no reduction in the water band \citep[][]{yang2015}. This has been interpreted as an indication for high-altitude haze or cloud layers in mid-L-type brown dwarfs, resulting in very small water opacities above the modulation-inducing particles. Furthermore, up to now, L-type brown dwarfs have only been found with low-amplitude variations.

Interestingly, the dusty L6-type W0047 shows both high-amplitude variations (typical to L/T dwarfs) and a very weak reduction in its amplitude in the 1.4~$\mu m$ water band (typical to L-dwarfs). Furthermore, W0047 is also unusual due to its extremely red \textit{[J-H]} color, which has been attributed to the presence of high-altitude particles (condensates or haze, \citealt[][]{hiranaka2016}). We followed the \citet{hiranaka2016} to model the wavelength-dependent amplitude by introducing a dust layer causing extinction, as described with Mie scattering-calculated opacities from Fig 5 of \citet{hiranaka2016}. We found that the model predictions are consistent with the observations. The model predicts that the observed variations could be explained by a change of particle column density of the order of $10^{7} cm^{-2}$ and sub-micron-sized grains (effective grain radius of 0.3$-$0.4 $\mu m$, see Eq (2) in \citealt{hiranaka2016} for definition of effective radius). Smaller effective grain sizes (e.g., 0.1, 0.2 $\mu m$) give smaller extinction coefficients, resulting in lower amplitudes,  while larger grain sizes (1 $\mu m$) give higher extinction coefficients, translating to much higher amplitudes and weaker dependence on the wavelengths.  The grain size is consistent with the grain size estimated in \citet{marocco2014} to explain the unusually red spectrum of W0047 with dust extinction caused by corundrum ($\mathrm{Al_2O_3})$ and enstatite ($\mathrm{MgSiO_3}$). Furthermore, based on the results of \citet{hiranaka2016}, the typical column number density of red L dwarfs' haze particles is on the order of $10^{8} cm^{-2}$ (see Fig. 12 of \citealt{hiranaka2016}), so our estimated change of haze particle column density is on the order of ten percent compared to total haze particle number column density in L dwarf's atmosphere. The actual change of column density can be larger if it occurs only in a small area (e.g., spots) on brown dwarfs. Future observations at longer wavelengths, the K-band and beyond, will provide stronger constraints on the extinction coefficients and grain sizes.

\section{Summary}
We present time-resolved near-infrared HST/WFC3 observations of the extremely red L6.5-type 20~M$_{Jup}$ intermediate-gravity brown dwarf as the first results from the HST Large Treasury program {\em Cloud Atlas}. The key results of our study are as follows:
\begin{enumerate}
\item We discovered large-amplitude rotational modulations in W0047's near-infrared light curve, making it the brown dwarf with the third-largest known near-infrared amplitude variations, after the T2 dwarf 2M2139 and T0 dwarf Luhman 16B. Our target is also the most variable L dwarf currently known.
\item  The fact that W0047 rotational modulation amplitude is almost an order of magnitude larger than other known variable L-dwarfs and is also one of the reddest known L-dwarfs suggests a possible connection between dusty atmospheres and high-amplitude variability. We used \citet{hiranaka2016} extinction coefficient to explain the wavelength-dependent amplitude with haze particle size of around 0.4 $\mu m$ and change of haze particle column density around $10^7cm^{-2}$, on the order of ten percent of total dust column density of L dwarfs estimated by \citet{hiranaka2016}.

\item Assuming that the light curve is single peaked and well approximated by a sine wave, the rotation period of W0047 is 13.20$\pm$0.14~hour, similar to that of Saturn and to the exoplanets $\beta$~Pic~b and 2M1207b. Our period estimation also serves as lower limit for more complex lightcurve profile.
\end{enumerate}
Our discovery demonstrates that high amplitude variable brown dwarfs are not limited to the L/T transition, suggesting that dusty atmospheres are more variable, and provides an outstanding benchmark target for cloud structure studies for ground- and space-based telescopes, such as the \textit{Hubble} and \textit{James Webb Space Telescope}.
 
\section*{Acknowledgement}
We thank the anonymous referee for useful comments that improved the manuscript. We would like to thank Min Fang, Alex Bixel, Kevin Wagner, Carol Chia-Jung Yang and Belle Yu-Ya Huang for providing useful comments. Ben W. P. Lew is supported in part by the Technology Research Initiative Fund (TRIF) Imaging Fellowship, University of Arizona. 
Support for Program number 14241 was provided by NASA through a grant from the Space Telescope Science Institute, which is operated by the Association of Universities for Research in Astronomy, Incorporated, under NASA contract NAS5-26555
Based on observations made with the NASA/ESA Hubble Space Telescope, obtained in GO program 14241 at the Space Telescope Science Institute. 
This publication makes use of data products from the Two Micron All Sky Survey, which is a joint project of the University of Massachusetts and the Infrared Processing and Analysis Center/California Institute of Technology, funded by the National Aeronautics and Space Administration and the National Science Foundation.


\end{document}